\documentclass[12pt]{article}
\usepackage{epsfig}
\usepackage{amsfonts}
\begin{document}
\begin{titlepage}
\begin{center}

{\Large Superstatistics in high energy physics: Application to
cosmic ray energy spectra and $e^+e^-$ annihilation}

\vspace{2.cm} {\bf Christian Beck}

\vspace{2.cm}

School of Mathematical Sciences, Queen Mary University of London,
Mile End Road, London E1 4NS, UK.

\vspace{5cm}

\end{center}

\abstract{We work out a superstatistical description
of high-energy scattering processes
that takes into account
temperature fluctuations in small volume elements.
For $\Gamma$-distributed
fluctuations of the inverse temperature one effectively obtains
formulas similar to those used in nonextensive statistical mechanics, whereas
for other temperature distributions more general superstatistical models arise.
We consider two main examples:
Scattering processes of cosmic ray particles and
$e^+e^-$ annihilation processes. In both cases one obtains
excellent fits of experimentally measured energy spectra and cross sections.}

\vspace{1.3cm}

\end{titlepage}

\section{Introduction}

Superstatistical techniques have been recently successfully applied to a
large variety of complex systems, for example
hydrodynamic
turbulence \cite{prl01,reynolds,BCS,prl07}, defect
turbulence \cite{daniels},
share price dynamics
\cite{ausloos,straetennew}, random matrix theory
\cite{RMT,abul-magd2}, random networks
\cite{abe-thurner},
wind velocity fluctuations \cite{rapisarda,kantz},
hydro-climatic fluctuations \cite{porpo}, the statistics of train
departure delays \cite{briggs} and models of the metastatic
cascade in cancerous systems \cite{chen}.
The basic idea underlying this approach is that there is
an intensive parameter, for example the inverse temperature $\beta$
or the energy dissipation
in turbulent systems,
that exhibits fluctuations on a large time
scale (large as compared to internal relaxation times of the
system under consideration). As a consequence, one can model these
types of complex systems by a kind of superposition of ordinary statistical
mechanics with varying temperature parameters, in short a superstatistics
\cite{beck-cohen,supergen,touchette,chavanis,abc,jizba,queiros,straeten}.
The stationary distributions of superstatistical systems
deviate from ordinary Boltzmann-Gibbs statistical mechanics
and can exhibit asymptotic power laws, stretched exponentials,
or other functional forms in the energy $E$
\cite{touchette}.

In this paper we work out potential applications
of this concept
in high energy physics. In scattering processes at high energies,
the effective interaction volume, as well as the number of
particles involved, can be rather small, and hence temperature fluctuations
can play a very important role \cite{cosmic,wilk04,wilknew}. These can either be temperature
fluctuations from scattering event to scattering event, i.e.\
temporal fluctuations, or they may be related a nonequilibrium
situtations where different spatial regions have different temperature,
i.e.\ spatial fluctuations.

Our starting point for a suitable thermodynamic
model will be Hagedorn's theory \cite{hage,hage2},
which yields a statistical
description of a selfsimilar `fireball' of particles produced
in scattering events.
Hagedorn's theory models the hadronization
cascade
from a statistical mechanics point of view (of course QCD was
not known at the time when he wrote the seminal paper \cite{hage}). His theory
is regularly in use to describe
heavy ion collisions \cite{ion1,ion,ion2}.
The
basic assumption is that in the scattering region
the density of states grows so rapidly that
the effective temperature cannot exceed a certain maximum temperature, the
Hagedorn temperature $T_H$ \cite{CERN}. The value is approximately $T_H\approx 180$ MeV
and it describes the confinement phase transition.
The Hagedorn phase transition is also
of
fundamental interest
in string theories \cite{string1, string2, string3}.


In this paper we will consider a superstatistical extension of the
Hagedorn theory, starting from a $q$-generalized version previously introduced
in \cite{beck00}. The predictions of this
generalized statistical mechanics model are
in excellent agreement with measured experimental
data. We will illustrate this for two main examples:
i) observed energy spectra of cosmic rays \cite{cosmic} ii)
experimentally measured cross
sections in $e^+e^-$ annihilation \cite{beck00}.


The Hagedorn theory of scattering processes
is known to give correct predictions of cross sections
and energy spectra for center of mass energies $E_{CMS}<10
GeV$, whereas for larger energies there is experimental evidence
from various collision experiments that power-law behaviour of
differential cross sections sets in.
This power-law is not contained in the original Hagedorn theory
but can be formally obtained if one extends
the original Hagedorn theory to a superstatistical one, along
the lines sketched in this paper.
In fact, if fluctuations are taken into account then
power laws can arise in quite a natural way
in various ways, and often lead to
Tsallis type of generalized distribution functions\cite{tsa1,wilk,quarati,prl01,beck-cohen}.
While in this paper
we concentrate on cosmic ray statistics and transverse
momentum spectra in
$e^+e^-$ annihilation, it is quite clear that
related superstatistical
techniques can be applied to
other scattering data as well,
for example heavy ion collisions
\cite{ion2, biro},
$pp$ collisions \cite{gorenstein} and $p\bar{p}$ collisions
\cite{wilknew}.


\section{Superstatistical modelling of temperature fluctuations in scattering experiments}

In the superstatistics approach \cite{beck-cohen}
one assumes that locally the system under consideration reaches local equilibrium but
on a large spatio-temporal scale there are temperature fluctuations,
described by a probability density $f(\beta)$ which describes
the probability to observe a certain inverse temperature $\beta$ in
a given spatial area.
One thus gets a kind of mixing (or superposition)
of many equilibrium distributions which effectively decribe
the driven nonequilibrium system with a stationary state (see \cite{sigmaphi08} for a recent
review discussing various applications).

For high energy scattering processes, it is clear that
the larger the center of mass energy $E_{CMS}$ of the
collision process is, the smaller is the volume $r^3$ probed, due to the
uncertainty relation $\frac{1}{c} E_{CMS} \cdot r \sim O(\hbar)$.
This means, the effective interaction volume where a thermodynamic
description of the collision process makes sense will become
smaller and smaller with increasing $E_{CMS}$. However, a smaller
volume means larger temperature fluctuations. This is in particular
relevant if we
repeat our scattering experiment several times or if
we have different scattering events in different spatial regions
which all contribute to our data.
It thus makes sense
to consider at large energies $E_{CMS}$ a superstatistical description
of scattering events which takes into account
local temperature fluctuations.

Assume that locally
some value of the fluctuating inverse temperature $\beta$
is given. We then expect
the momentum of a randomly picked particle in this region to be
distributed according to the relativistic Maxwell-Boltzmann
distribution
\begin{equation}
p(E|\beta)=\frac{1}{Z(\beta)}E^2 e^{-\beta E}. \label{max}
\end{equation}
Here $p(E|\beta)$ denotes the conditional probability to observe
a particle with energy $E=\sqrt{\vec{p}^2c^2+m^2c^4}$, given
some value of $\beta$. For highly relativistic
particles we can neglect the rest mass $m$ so that
$E=c|\vec{p}|$, where $\vec{p}$ is the momentum.
The normalization constant is given by
\begin{equation}
Z(\beta)=\int_0^\infty E^2 e^{-\beta E} dE=\frac{2}{\beta^3} .
\end{equation}
Now let us take into account local temperature fluctuations in the
small interaction volumes where scattered particles are produced.
We have to consider some suitable probability density $f(\beta)$
of the inverse temperature in the various interaction volumes.
In a long series of experiments we will then observe the marginal distribution
obtained by integrating over all $\beta$
\begin{equation}
p(E)=\int_0^\infty p(E|\beta)f(\beta)d\beta .\label{9}
\end{equation}
These types of distributions are generally studied in superstatistical
models, and are known to exhibit fat tails whose asymptotic
decay with $E$ depends on how $f(\beta)$ behaves for $\beta \to 0$
(more details on this in \cite{touchette}). Of course, the validity of the
superstatistical modelling approach requires sufficient time scale separation
so that the system relaxes to local equilibrium fast enough before the next
temperature fluctuation takes place
\cite{BCS, straetennew, abc, straeten}.

It has been argued in previous work \cite{BCS} that there are basically
three different distributions $f(\beta)$ that are relevant for large
classes of complex systems: These are the $\chi^2$-distribution,
the inverse $\chi^2$ distribution, and the lognormal distribution.
Lognormal superstatistics is often observed in hydrodynamic
turbulence, due to the multiplicative nature of the Richardson
cascade \cite{BCS,prl07}. Inverse $\chi^2$ superstatistics
plays an important role in random matrix theory
\cite{RMT, abul-magd2},
as well as in medical statistics \cite{chen}. For
high energy physics, $\chi^2$ superstatistics
is most relevant, although
other superstatistics may play a role as well (for example, one could think
about a superstatistics generating Kaniadakis statistics \cite{kaniadakis}).

For $\chi^2$ superstatistics (or equivalently $\Gamma$ superstatistics), the distribution $f(\beta)$
is given by
the $\chi^2$-distribution of degree $n$, i.e.
\begin{equation}
f (\beta) = \frac{1}{\Gamma \left( \frac{n}{2} \right)} \left\{
\frac{n}{2\beta_0}\right\}^{\frac{n}{2}} \beta^{\frac{n}{2}-1}
\exp\left\{-\frac{n\beta}{2\beta_0} \right\}. \label{fluc}
\end{equation}
The $\chi^2$-distribution is a typical distribution that naturally
arises in many circumstances, for example if $n$
independent Gaussian random variables $X_i,\;i=1,\ldots ,n$ with
average $0$ and the same variance are squared and added. If we write
\begin{equation}
\beta:=\sum_{i=1}^{n} X_i^2 \label{Gauss}
\end{equation}
then $\beta$ has the probability density function (\ref{fluc}).
The average of the fluctuating $\beta$ is given by
\begin{equation}
\langle \beta \rangle =n\langle X_i^2\rangle=\int_0^\infty\beta f(\beta) d\beta= \beta_0
\end{equation}
and the variance by
\begin{equation}
\langle \beta^2 \rangle -\beta_0^2= \frac{2}{n} \beta_0^2.
\end{equation}
The integral (\ref{9}) with $f(\beta)$ given
by (\ref{fluc}) and $p(E|\beta)$ given by (\ref{max})
is easily evaluated and one obtains
\begin{equation}
p(E) \sim \frac{E^2}{( 1+b(q-1)E)^{\frac{1}{q-1}}}
\label{14}
\end{equation}
where
\begin{equation}
q=1+\frac{2}{n+6} \label{qwert}
\end{equation}
and
\begin{equation}
b=\frac{\beta_0}{4-3q}.
\end{equation}
Note that the partition function
$Z(\beta)$ entering into eq.~(\ref{9})
is  $\beta$-dependent. Different $\beta$-dependencies
of the partition
function $Z(\beta)$ lead to different answers if the integration
over $\beta$ is performed. In particular, the precise relation between
$q$ and $n$ depends on this.
This was for the first time correctly worked
out in \cite{prl01}.

The distribution (\ref{14}) is a $q$-generalized relativistic
Maxwell-Boltzmann distribution in the formalism of nonextensive
statistical mechanics \cite{tsa1}. These kind of distributions can be directly
obtained by maximizing the $q$-entropies \cite{tsa1}
\begin{equation}
S_q=k \frac{1}{q-1} (1-\sum_i p_i^q )
\end{equation}
and multiplying with the available phase space volume.
The $p_i$ are the probabilities of the
microstates $i$. The $q$-entropies contain the Shannon entropy
$S_1=-k\sum_i p_i \ln p_i$ underlying ordinary statistical
mechanics as a special case for $q=1$. Whereas $q=1$ corresponds to
the usual canonical ensemble with constant temperature,
the
Tsallis-canonical ensemble obtained for $q>1$ is capable of
describing temperature fluctuations,
assuming that $\beta$ is $\chi^2$-distributed.
We thus have a plausible physical mechanism why a Tsallis-like
statistical description makes sense if a
suitable intensive parameter fluctuates \cite{prl01,wilk}. Tsallis statistics with
$q=3$ (respectively $q=-1$ if the escort formalism \cite{BS, mendes}
is used) also plays an important role for chaotically quantized
scalar fields \cite{nonli,book}, dark energy models \cite{prd} and
moduli field dynamics \cite{erice}.

\section{Superstatistical partition functions}

When experimental scattering data are collected, one often looks at
momentum distributions of a particular particle that are produced
by repeating the same experiment many times. In each scattering event,
the effective temperature (given by the heat bath of surrounding
particles) will fluctuate from event to event.
Since the scattering experiment is repeated many times in an
independent way, and since we concentrate on the
statistics of just one particle rather then many-particle states,
it makes sense to directly integrate out the
temperature fluctuations and to consider effective
1-particle superstatistical Boltzmann
factors.

Let us consider
particles of different types and label the particle types
by an index $j$. Each particle can be in a certain momentum state labelled
by the index $i$.
The energy associated with this state is
\begin{equation}
\epsilon_{ij}=\sqrt{{\vec{p}}^2_i+m_j^2},
\end{equation}
using units where $c=1$.
We may now define an effective 1-particle Boltzmann factor by\footnote{To describe 2-particle states with
the same temperature fluctuations surrounding both
particles one has
to proceed in a slightly different way, see \cite{prl01}.}

\begin{equation}
x_{ij} := \int_0^\infty d\beta \, f(\beta) e^{-\beta \epsilon_{ij}}.
\end{equation}
For example, for Tsallis statistics $f(\beta)$ is a $\chi^2$ distribution and one has
\begin{equation}
x_{ij}=(1+(q-1)b\epsilon_{ij})^{-\frac{1}{q-1}},
\end{equation}
where $b^{-1}$ is proportional to the average temperature, and $q-1$
is a measure of the strength of inverse temperature fluctuations.
The effective Boltzmann factor $x_{ij}$
approaches the ordinary Boltzmann factor $e^{-b\epsilon_{ij}}$
for $q\to 1$.

The simplest way to introduce a generalized grand
canonical partition function (which, as said before, regards all
temperature fluctuations to be independent) is as follows:
\begin{equation}
Z=\sum_{(\nu)} \prod_{ij}x_{ij}^{\nu_{ij}}
\end{equation}
Here $\nu_{ij}$ denotes the number of particles of type $j$ in
momentum state $i$. The sum
$\sum_{(\nu)}$ stands for a summation over all possible particle numbers.
For bosons one has $\nu_{ij}=0,1,2,\ldots ,\infty$,
whereas for fermions one has $\nu_{ij}=0,1$.
It follows that
for bosons
\begin{equation}
\sum_{\nu_{ij}} x_{ij}^{\nu_{ij}}=\frac{1}{1-x_{ij}} \;\;\;\;\;\;\; (bosons)
\end{equation}
whereas for fermions
\begin{equation}
\sum_{\nu_{ij}} x_{ij}^{\nu_{ij}}=1+x_{ij} \;\;\;\;\;\;\;(fermions)
\end{equation}
Hence the partition function can be written as
\begin{equation}
Z=\prod_{ij}\frac{1}{1-x_{ij}}\prod_{i'j'}(1+x_{i'j'}),
\end{equation}
where the prime labels fermionic particles. For the logarithm
of the partition function we obtain
\begin{equation}
\log Z= -\sum_{ij} \log(1-x_{ij})+\sum_{i'j'}\log(1+x_{i'j'}).
\end{equation}
One may actually proceed to continuous variables
by replacing
\begin{equation}
\sum_i [...] \to \int_0^\infty \frac{V_0 4 \pi p^2}{h^3}[...]dp
=\frac{V_0}{2\pi^2} \int_0^\infty p^2[...]dp \;\;\;\;(\hbar=1)
\end{equation}
($V_0$: volume of the interaction region) and
\begin{equation}
\sum_j [...] \to \int_0^\infty \rho(m)[...]dm,
\end{equation}
where $\rho(m)$ is the mass spectrum.

Let us now formally calculate the average occupation number of a
particle of species $j$ in the momentum state $i$. We obtain
\begin{equation}
\bar{\nu_{ij}} =x_{ij} \frac{\partial}{\partial x_{ij}} \log Z
= \frac{x_{ij}}{1\pm x_{ij}} .
\end{equation}
For example, for the case of Tsallis statistics one obtains
\begin{equation}
\bar{\nu_{ij}}
= \frac{1}{(1+(q-1)b\epsilon_{ij})^{\frac{1}{q-1}}\pm 1}
\end{equation}
where the $-$ sign is for bosons and the $+$ sign for fermions.

In order to single out a particular particle of mass $m_0$,
one can formally work with the mass spectrum $\rho(m)=\delta(m-m_0)$.
To obtain the probability to observe a particle of mass $m_0$ in
a certain momentum state, we have to multiply
the average occupation number with the available volume in momentum
space. An infinitesimal volume in momentum space can be written as
\begin{equation}
dp_xdp_ydp_z=dp_Lp_T\sin\theta dp_T d\theta
\end{equation} where $p_T=\sqrt{p_y^2+p_z^2}$ is the transverse monentum and
$p_x=p_L$ is the longitutinal one.  By integrating over all $\theta$ and $p_L$
one finally arrives at a
probability density $w(p_T)$ of transverse momenta given by \begin{equation}
w(p_T) =const \cdot p_T \int_0^\infty dp_L
\frac{1}{(1+(q-1)b\sqrt{p_T^2+p_L^2+m_0^2})^{\frac{1  }{q-1}}\pm 1}.
\end{equation}
Since the Hagedorn temperature is rather small (of the order of the $\pi$ mass),
and $b^{-1}$ is of the order of the Hagedorn temperature,
under normal circumstances one has $b \sqrt{p_T^2+p_L^2+m_0^2} >>1$,
and hence the $\pm 1$ can be neglected if $q$ is close
to 1. One thus obtains for both
fermions and bosons the statistics
\begin{equation}
w(p_T)\approx const \cdot p_T \int_0^\infty dp_L
\left( 1+(q-1)b \sqrt{p_T^2+p_L^2+m_0^2}\right)^{-\frac{1}{q-1}} \label{28}
\end{equation}
which, if our model assumptions are satisfied,
should determine the $p_T$ dependence of experimentally
measured particle spectra. The differential
cross section $\sigma^{-1} d\sigma /dp_T$ is expected to be
proportional to $w(p_T)$.

In nonextensive statistical mechanics, it is sometimes of advantage
to consider normalized $q$-expectation values \cite{mendes}, thus
implementing a formalism that is based on so-called escort distributions \cite{BS}.
If the escort formalism
is used then the power $\frac{1}{q-1}$
in the above formula is
replaced by $\frac{q}{q-1}$, a simple reparametrization. For more general classes
of superstatistics described by general $f(\beta)$
the $q$-exponential is replaced by $\int_0^\infty f(\beta) \exp \{-\beta \sqrt{p_T^2+p_L^2+m_0^2}\} d\beta$.

\section{Comparison with measured cosmic ray energy spectra}

\begin{figure}
\epsfig{file=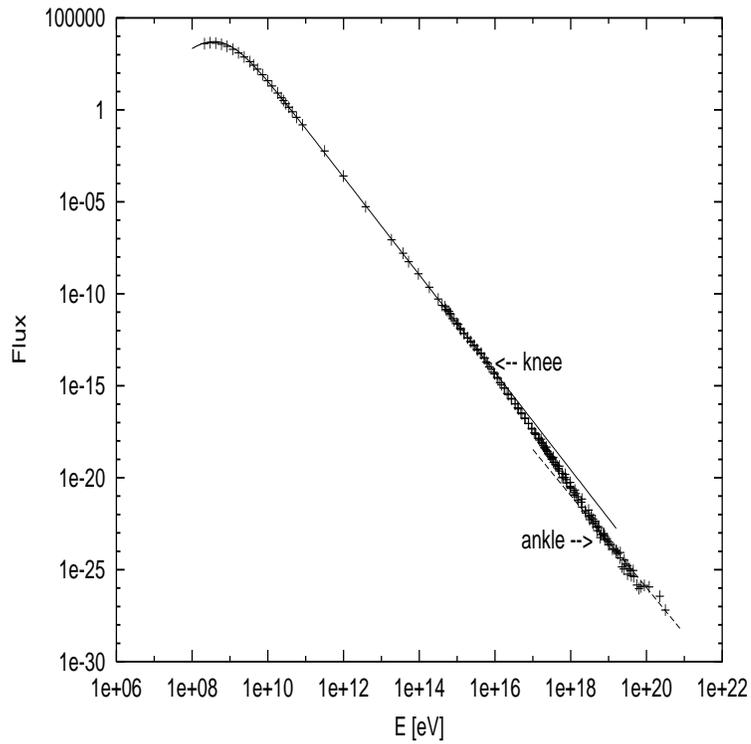, width=10cm, height=10cm}
\caption{Measured energy spectrum of primary cosmic rays (in
units of $m^{-3}s^{-1}sr^{-1} GeV^{-1}$)
 \cite{cosmic}. The solid line is the
formula (\ref{14}) with $q=1.215$, $b^{-1}=kT_0=107$ MeV and $C=5\cdot
10^{-13}$ in the above units. The dashed line is eq.~(\ref{14}) with $q=11/9$,
$kT_0=107$ MeV and $C$ smaller by a factor $1/50$.}
\end{figure}

Let us now look at measured data. Fig.~1 shows the
experimentally measured energy spectrum of
primary cosmic ray particles as observed on the earth \cite{chicago,pdg,high1,high2,high3,high4}.
The measured spectrum is very well fitted
over many decades of different energies by
the distribution (\ref{14}) \cite{cosmic} or other generalized nonextensive
distributions \cite{cos1}. The best fit is
obtained if the entropic index $q$ is chosen as
\begin{equation}
q=1.215
\end{equation}
and if the effective temperature parameter is given by
\begin{equation}
kT_0=b^{-1}=107 \; \mbox{MeV}.
\end{equation}

Let us now predict a plausible value of $q$ using the
superstatistical approach.
The variables $X_i$ in eq.~(\ref{Gauss}) describe the independent degrees of freedom
contributing to the fluctuating inverse temperature $\beta$. At very large center
of mass energies $E_{CMS} \to \infty$, the interaction region is very small, and all
relevant degrees of freedom are basically represented by the 3
spatial dimensions into which heat can flow. We may physically
interpret $X_i^2$ as the heat loss in the spatial $i$-direction,
$i=x,y,z$, during the collision process that generates the primary cosmic
ray particle. The more heat is lost, the smaller is the local
$kT$, i.e. the larger is the local $\beta$ given by (\ref{Gauss}).
The 3 spatial degrees of freedom yield $n=3$ as the smallest
possible value of $n$ or, according to (\ref{qwert}),
\begin{equation}
q=\frac{11}{9}=1.222. \label{qmax}
\end{equation}
For cosmic rays $E_{CMS}$ is very large, hence we expect a
$q$-value that is close to this asymptotic value. The fit in
Fig.~1 in fact uses $q=1.215$, which agrees with the predicted
value in eq.~(\ref{qmax}) to about 3 digits
(similar $q$-values were also obtained
in \cite{cos1}).

For smaller center of mass energies, according
to $\frac{1}{c} E_{CMS} \cdot r \sim O(\hbar)$,  the interaction region will
be bigger and more effective degrees of freedom within this bigger
interaction region will contribute to the fluctuating temperature.
Hence we expect that for smaller $E_{CMS}$ $n$ will be larger than
3, or $q$ will be smaller than $11/9$.

Finally, for  $E_{CMS} \to 0$, $q\to 1$ and ordinary
statistical mechanics is recovered. In this classical limit case,
the relevant interaction region $r^3$ where a thermodynamic
description makes sense becomes very large, and within this large
region a large number of independent degrees of freedom $n$
contributes to the fluctuating temperature, represented by many
different particles. According to eq.~(\ref{qwert}), the limit
$n\to \infty$ is equivalent to $q\to 1$ and in this limit the
$\chi^2$-distribution degenerates to a delta function $\delta
(\beta -\beta_0)$.

It is reasonable to assume \cite{pdg}
that the `knee' at $E\approx 10^{16}$ eV is due to the
fact that one has reached the maximum energy scale to which
typical galactic accelerators can accelerate. This then
implies a rapid fall in the number of observed events with a
higher energy, i.e. a steeper slope in Fig.~1 between about
$10^{16}$ and $10^{19}$ eV. The `ankle' at $E\approx 10^{19}$ eV
may then be due to the fact that a higher energy population of
cosmic ray particles takes over from a lower energy population.
This higher energy
population may have
a different origin (for example,
extragalactic origin). The new population has a significantly smaller flux
rate but can reach much larger energies. As a matter of fact the cosmic
accelerators underlying the production process of this new species of
cosmic rays must have a much larger center of mass energy
$E_{CMS}$ than the ankle energy $\sim 10^{19}$ eV, so $q$ should
be given by its asymptotic value $11/9$, whereas
the effective temperature $T_0$ should be the same as before. The
dashed line in Fig.~1 corresponds to our formula with $q=11/9$,
$k\tilde{T}=107$ MeV and a flux rate that is smaller by a factor $1/50$ as
compared to the high-flux generation of cosmic rays. This is
consistent with the data.

\begin{figure}
\epsfig{file=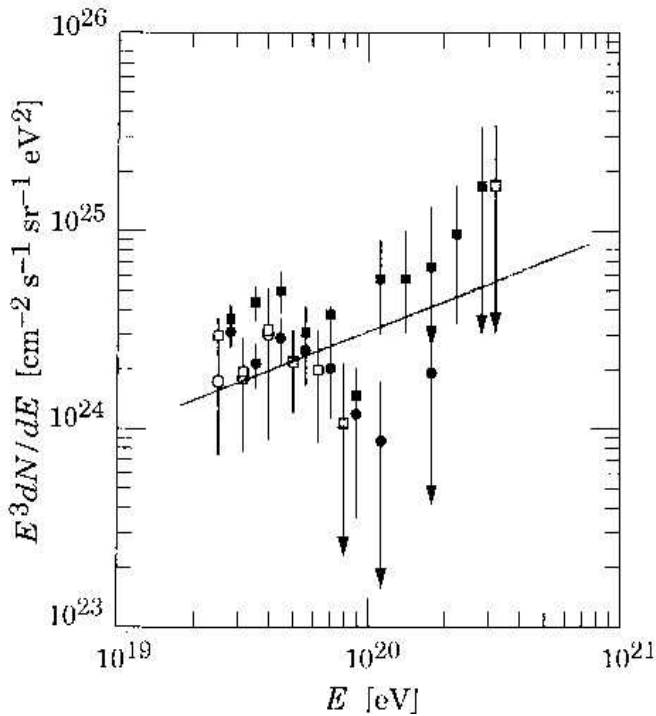, width=10cm, height=10cm}
\caption{Measured cosmic ray energy spectrum $E^3 \cdot dN/dE$
at largest
energies (data from \cite{pdg,high1,high2,high3}). The straight line is
a power law with exponent $\alpha =5/2$ (corresponding
to $q=11/9$).}
\end{figure}

Formula (\ref{14}) predicts asymptotic power-law behavior of the
measured energy spectrum. For large $E$ one has $p(E)\sim
E^{-\alpha}$ where the index $\alpha$ is given by
\begin{equation}
\alpha =\frac{1}{q-1}-2.
\end{equation}
$q=1.215$ implies $\alpha =2.65$ (for moderately large energies),
whereas for largest energies the asymptotic value $q=11/9$ implies
$\alpha =5/2$. As shown in Fig.~2, the largest-energy events are
compatible with such an asymptotic power law exponent.


\section{Comparison with experimentally measured
differential cross sections in $e^+e^-$ annihilation}

\begin{figure}
\epsfig{file=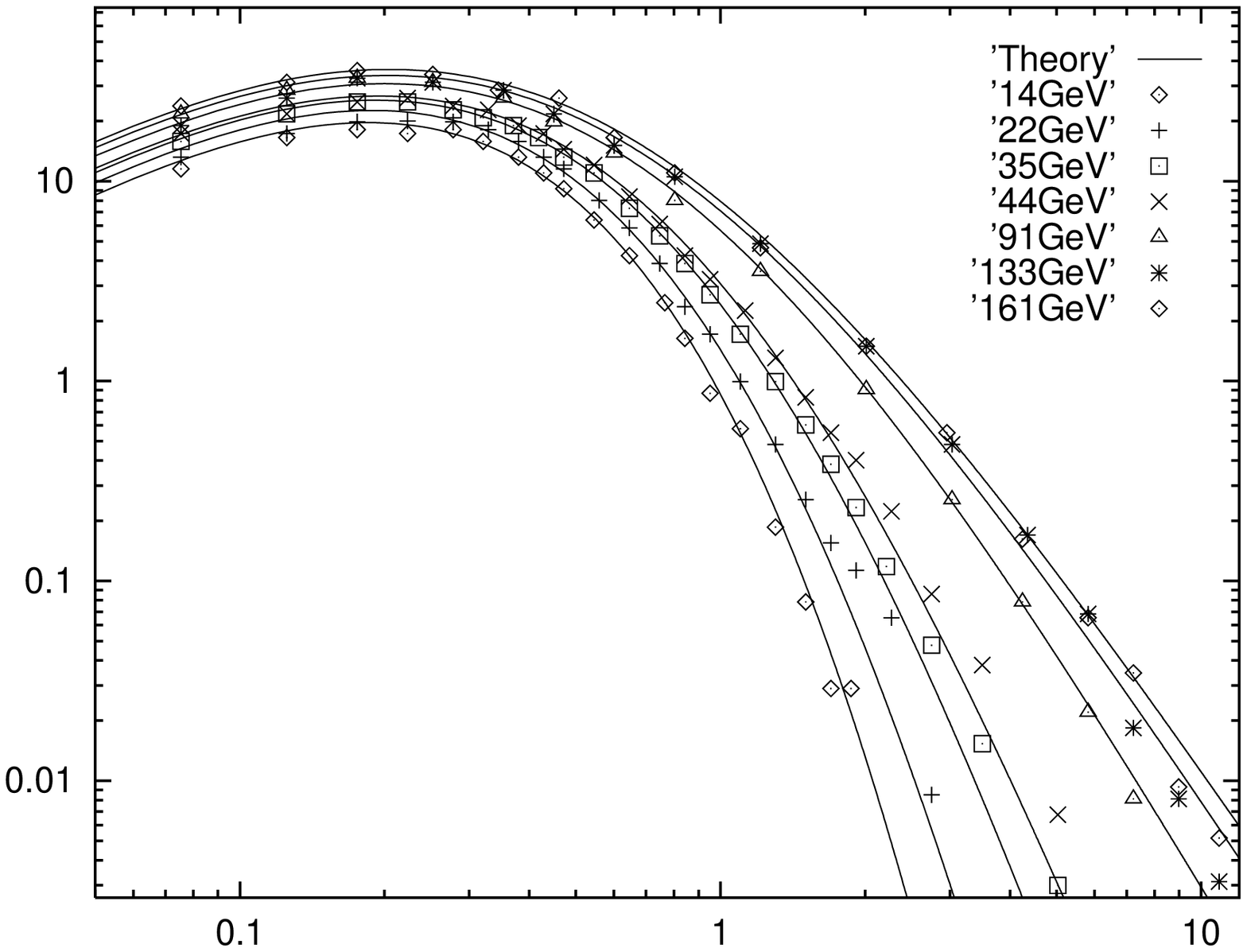, height=4.6in, width=4.6in,angle=-90}
\caption{
Differential cross section as a function of the
transverse momentum $p_T$ for various center of mass energies $E$.
The data correspond to measurements of the TASSO ($E\leq 44$ GeV)
and DELPHI ($E\geq 91$ GeV)
collaboration. The solid lines are given by
the analytic formula (\ref{here7}).}
\end{figure}
Let us now proceed to our second example, differential cross sections for
transverse momenta in $e^+e^-$ annihilation experiments \cite{beck00,bediaga}. If one
uses the escort formalism \cite{BS, mendes} then
formula (\ref{28})
implies

\begin{equation}
\frac{1}{\sigma} \frac{d\sigma}{dp_T} = C u \int_0^\infty
dx \; \left( 1+(q-1)\sqrt{x^2+u^2+m_\beta^2} \right)^{-\frac{q}{q-1}}
\label{exact} \end{equation}
Here $x=p_L/T_0$, $u=p_T/T_0$ and $m_\beta:=m_0/T_0$ are the
longitudinal momentum, transverse momentum and mass in units of
a temperature parameter $T_0$ that is of the same order of
magnitude as the Hagedorn temperature.
$C$ is a suitable
constant related to multiplicity.

Generally, one knows that the interaction energy in a nonextensive
system increases with increasing entropic index $q$. Since this energy must be taken
from somewhere, it is most natural to assume that it is taken from
the heat bath. So the average temperature parameter $T_0$
should slighly decrease with increasing $q$. In \cite{beck00} a linear dependence of $T_0$ on $q$
was postulated, of the simple form
\begin{equation}
T_0=\left( 1-\frac{q}{3} \right) T_H \label{tqq}
\end{equation}
where $T_H=180 \; MeV$ is the Hagedorn temperature.

We also have to clarify the dependence of $q$ on $E_{CMS}$. Clearly, for
$E_{CMS} \to \infty$ our argument in section 2 and 4 suggests the value
$q=11/9$, whereas for $E_{CMS} \to 0$ one has $q=1$, i.e the ordinary
Hagedorn theory without fluctuations. In \cite{beck00}, a smooth interpolation
between these values was suggested, of the form
\begin{equation}
q(E_{CMS})=\frac{11-e^{-E_{CMS}/E_0}}{9+e^{-E_{CMS}/E_0}} \label{here10}
\end{equation}
where $E_0 \approx 45.6 \; GeV$ is about half of the $Z_0$ mass.

Finally, one also needs to know the multiplicity (the
average number of produced charged particles) in order to
estimate the cross section.
The multiplicity $M$ as a function of $E_{CMS}$ has been independently
measured in many experiments\cite{passon}. A good fit of the
experimental data in the relevant energy region
is the formula \cite{beck00}
\begin{equation}
M=\left( \frac{E_{CMS}}{T_0^{q=1}}\right)^{5/11}
\label{here88}
\end{equation}
The final formula derived for the cross section in \cite{beck00} involves
a further approximation step to perform the integration over $x$ and is then finally given by
\begin{equation}
\frac{1}{\sigma}\frac{d\sigma}{dp_T}= \frac{1}{T_0} M p(u) \label{here7}.
\end{equation}
where $p(u)$ is the normalized probability density
\begin{equation}
p(u)=\frac{1}{Z_q} \;
u^{3/2} \left( 1+(q-1)u \right)^{-\frac{q}{q-1}+\frac{1}{2}}
\label{here3}
\end{equation}
with normalization constant
\begin{equation}
Z_q=(q-1)^{-5/2} B\left( \frac{5}{2}, \frac{q}{q-1}-3 \right)  .
\end{equation}

Formula (\ref{here7}) with $p(u)$ given by
eq.~(\ref{here3}), $q(E_{CMS})$ given by eq.~(\ref{here10}),
$T_0(q)$ given by eq.(\ref{tqq})
and multiplicity $M(E_{CMS})$ given by eq.~(\ref{here88})
turns out to very well reproduce
the experimental results of measured transversal cross sections
for all energies $E_{CMS}$. This is illustrated
in Fig.~3 which shows the measured differential cross section versus $p_T$. The solid lines are given by formula (\ref{here7}) for the various
center of mass energies $E_{CMS}$.
One obtains excellent agreement with the measured data
of the
TASSO and DELPHI collaborations \cite{TASSO, DELPHI}.
Remarkably, for the largest
energies the best fitting parameter $q$ is again given by $q\approx 11/9$,
in agreement
with the data for the cosmic rays. Moreover, at the largest energies,
according to eq.~(\ref{tqq}) $T_0$ is given by $107 \; MeV$, again in agreement with what was used for the
cosmic ray data in Fig.~1. Our
superstatistical argument in section 2 and 4,
relating $q$ to temperature fluctuatuons and predicting from this $q\approx 11/9$
for largest energies $E_{CMS}$, is indeed generally applicable.

\section{Conclusion}

In this paper we have dealt with a superstatistical generalization
of the Hagedorn theory which is generally applicable to
describe the statistics of scattered particles
produced in high-energy collisions. Our approach is based on
taking into account temperature fluctuations in small interaction
volumes. For $\chi^2$-distributed inverse temperature one effectively
ends up with formulas that are similar to
those used in non-extensive versions of statistical mechanics.

At large energies the $\chi^2$-superstatistical approach implies that
energy spectra of particles and cross sections
decay with a power law. This power law is indeed
observed for various experimental data.
We obtained
formulas that are in very good agreement
with experimentally measured data for cosmic rays and
$e^+e^-$ annihilation. In particular, the superstatistical
approach allowed us to give a concrete prediction for $q$ at largest
center of mass energies, namely $q \approx 11/9$.
Generally,
it appears
that high-energy scattering data do not only yield valuable
information on elementary particle physics, but they may also
be regarded as test grounds to further develop
generalized versions of statistical mechanics.

\end{document}